\documentstyle[twoside,fleqn,espcrc2]{article}

\newcommand{\AmS}{{\protect\the\textfont2
  A\kern-.1667em\lower.5ex\hbox{M}\kern-.125emS}}

\title{The Crumpling Transition Revisited }

\author{M. Baig\address{Departament de F\'\i sica, Universitat
Aut\`onoma de Barcelona,
08193 Bellaterra, Spain.}
\thanks{Research supported by CICYT contract AEN90-0028}
        and
D. Espriu\address{D.E.C.M., Universitat de Barcelona,
Diagonal, 647, 08028 Barcelona, Spain.}
\thanks{Research supported by CICYT contract AEN90-0033 and
       CEE Science Twinning Grant SC1-0337-CA} }

\begin{document}

\begin{abstract}
The ``crumpling" transition, between rigid and
crumpled surfaces, has been object of much discussion over the past years.
The common lore is that such transition should be of second order.
However, some lattice versions of the rigidity term on
fixed connectivity surfaces seem
to suggest that the transition is of higher order instead. While some
models exhibit what appear to be lattice artifacts, others are really
indistiguishable from models where second order transitions have been
reported and yet appear to have third order transitions.
\end{abstract} 
\maketitle

\section{INTRODUCTION}

Recently
we have learnt a good deal about gravity coupled to conformal
matter with a central charge $c < 1$\cite{KPZdistKaz}.
However, after many
years of hard work we still do not understand very well the
behaviour of random surfaces in a more interesting range
of dimensions. Sometime ago\cite{polyakov} it was pointed out
that some of the difficulties encountered in quantizing two dimensional
gravity could be overcome
by adding to the area term (Nambu-Goto action) a piece proportional to the
extrinsic curvature of the surface. This term is called rigidity
and induces a correlation between the normals. In the continuum
there is a unique operator of the right dimension. It is
\begin{equation}
\nabla_a n^\mu \nabla_a n^\mu
\end{equation}
where $n^\mu$ is the normal vector to the surface. This operator is
accompanied by a coupling constant $\kappa$ whose renormalization
group equation is known both in mean field\cite{me}
and in the one-loop approximation\cite{others}. The theory
is ultraviolet-free in $1/\kappa$ implying that at short distances
$\kappa$
grows and one expects the surface to be more rigid. The
$\beta$-function found in \cite{me,others} has
no non-trivial zeroes, so the rigidity term is driven
to zero in the infrared and the surface becomes
highly convoluted over long distances.

This sort of behaviour has been observed in condensed matter physics
in a number of systems. The pertinent relevant observable there
is called the persistence length and it roughly corresponds to the
average distance where the surface
can be considered ``flat". However,
many such systems seem to have a sharp transition at some finite value
of $\kappa$ where the surface sharply changes from being ``crumpled"
(with an infinite Hausdorff dimension, $d_H=\infty$) to ``smooth"
($d_H\simeq 2$). At this point the persistence length increases
dramatically and the specific heat shows a peak. Theoretical
expectations\cite{david} and numerical simulations seem
to suggest that the transition may be of second order.

These results are clearly at odds with the results of
\cite{polyakov,me,others}.
This can probably be understood as follows. The calculation of the
$\beta$-function in \cite{david} assumes for the energy of the surface
a combination of the bending (or rigidity) energy plus some stretching
energy depending on two Lam\'e coefficients, as appropriate to crystalline
membranes. This, however, has little to do with the action used in
\cite{polyakov,me,others}, namely the area action (for
surfaces of fixed
connectivity) or  the more popular
lattice version of the Polyakov action
that includes a sum over triangulations\cite{dynamical} but keeps the
intrinsic
length of the triangles fixed. Both formulations
contain, amongst other things, a sum over the conformal mode which is
clearly absent in the Lam\'e approach. (Incidentally,
at least in the continuum limit, Nambu-Goto and Polyakov formulations
are equivalent even
at the quantum level\cite{tim}, not just classically.
Of course it would be
extremely important to understand this equivalence in lattice versions,
simulations being much easier with fixed connectivity surfaces.)

 Random surfaces seem to possess some non-universal
features. For instance, the critical exponents associated to surfaces
constructed on a hypercubic lattice\cite{frohlich}
 are certainly different from those
obtained in a triangulated model\cite{gross}. It is quite possible
that this non-universality is not accidental, but reflects some
intrinsic disease of two dimensional gravity, such as, for instance,
lack of renormalizability for $c>1$. There are several possible
ways to test this idea. One is, of course, to add copies of
Ising models so that $c>1$. In this conference
we have heard a number of contributions in this
direction. Another possibility is to perturbe the
action of Nambu-Goto strings and check for deviations. This provides
an additional motivation to pay special attention to the
crumpling transition.

\section{RIGID SURFACE ACTIONS}

We are thus confronted with a twofold task. On the one hand
we want to construct models that correspond to ``fluid"
surfaces and not to crystalline ones. It is clear that dynamically
triangulated surfaces are of this type. Surfaces with fixed
connectivity may or may not. The simplest test is to check whether
the $\beta$-function behaves for large values of the rigidity
parameter as is supposed to do (fig. 1). Indirectly, this can be
inferred from the existence or not of second order phase transtitions.

\begin{figure}[htb] \vspace{20mm}
\caption{Expected $\beta$-functions (for large values of $\kappa$) for
crystalline (A) and fluid (B) surfaces.} \label{fig:1}
\end{figure}

Secondly, we want to see whether all the possible models have
similar phase transitions. Should this not be the case it would be
extremely difficult to select one in front of the others, of course.
The natural consequence would be that some cut-off effects remain
in two-dimensional gravity.

We shall work only with triangulated
surfaces of fixed connectivity and will
comment only at the end about dynamical triangulation models.
We use a toroidal topology throughout and present results
only for three embedding dimensions. As ``area" term we can choose the area
itself or the
so called gaussian action\cite{gaussian} that consists of adding the lengths
of the edges of the triangles squared. As for the rigidity
term we have considered three different possibilities that we
will denote $S_1$, $S_2$ and $S_3$.

$S_1$ is a popular choice, both in dynamical and fixed connectivity
surfaces. It is
\begin{equation}
S_1=\sum_{ij}(1-\cos\theta_{ij})
\end{equation}
The sum extends to the angles formed by the normals to adjacent
triangles. An innocent-looking variation of this action
consists in expanding the cosinus and retaining only the first
term
\begin{equation}
S_2=\sum_{ij} \theta_{ij}^2/2
\end{equation}
Finally, we have also re-analyzed the action \cite{plb,npb}
\begin{equation}
S_3=\sum_i{1\over \Omega_i}(\sum_{j(i)}(x_j-x_i))^2
\end{equation}
Needless to say that $S_1$, $S_2$ and $S_3$ have the
same naive continuum limit, namely (1). It has been
pointed out that $S_3$ appears to have some lattice artifacts
that yield ``corrugated" surfaces in the smooth
phase\cite{harnish}, but for all we know is perfectly alright
in the crumpled phase and even around the phase transition. The
authors of \cite{npb} concluded that this action has a third
order crumpling transition.

We have not found any differences in the geometrical properties
of the typical surfaces generated by
$S_1$ and $S_2$. Figure 2 shows two snapshots of
a $16^2$ surface first in the crumpled phase and then in the
smooth one with an action area$+S_1$.
The most salient feature
of the surface in the crumpled phase is, of course,
the presence of the spikes. Without the presence of the
extrinsic curvature these make the theory sick. Spikes have
altogether dissapeared in the smooth phase. Figure 3
shows the equivalent snapshots but now with an action
gaussian$+S_2$. Since gaussian actions are known not to have
spikes the surface is certainly smoother than before in the
crumpled phase, pushing noticeably the crumpling
transition towards lower values of $\kappa$.

\begin{figure}[htb]  \vspace{30mm}
\caption{Typical surfaces. Area$+S_1$}
\label{fig:2}
\end{figure}
\begin{figure}[htb] \vspace{30mm}
\caption{Typical surfaces. Gaussian$+S_2$}
\label{fig:3}
\end{figure}

\section{NUMERICAL RESULTS}

We will report here only on the specific heat results from our Monte Carlo
simulations of the above actions.
One first observation is that
the equilibration and relaxation times are substantially longer for $S_1$
than for $S_2$ and $S_3$. In all three cases there are crumpling transitions
and around the critical point the relaxation times increase dramatically.
In fact, it becomes crucial to have very good statistics and perform the
binning in a systematic manner to get meaningful results. To compute
the specific heat the have grouped the runs of the larger systems  in
bins of $10^5$, $2\times
10^5$, etc. sweeps until the specific heat levels off. Typically this
happens
around $4\times 10^5$ sweeps. Then
we accumulate typically five or six of such bins to assess the statistical
errors. The rule of thumb is to run until the error bars around the phase
transition are comparable to those of values of $\kappa$ far from it.
We have also used other methods, such as that of Ferrenberg and
Swendsen (FS), but we have found much more difficult to assess the errors
there and we have
decided finally to present our results in a more old fashioned way.
FS has been, however, very helpful to fix the length of the runs to
guarantee good thermalization.

Figures 4 and 5 show the results for area$+S_1$ and area$+S_2$. The vertical
scale is the same in both cases to ease the comparison. Figure 6 exhibits
the specific heat plot for gauss$+S_2$. The shift in the location of the
critical point and the very moderate growth are very much similar to
area$+S_2$ and in sharp contrast to fig. 4. In fact, using a gaussian or
an area action seems to matter little for the crumpling transition. In
contrast fig. 4 that collects the results for $S_1$ definitely shows
a more genuine second order behaviour with a clear growth of the specific
heat.

\begin{figure}[htb]   \vspace{40mm}
\caption{Specific heat. Area$+S_1$}
\label{fig:4}
\end{figure}
\begin{figure}[htb]   \vspace{40mm}
\caption{Specific heat. Area$+S_2$}
\label{fig:5}
\end{figure}
\begin{figure}[htb]  \vspace{40mm}
\caption{Specific heat. Gaussian$+S_2$}
\label{fig:6}
\end{figure}
We have collected
in Table 1 our tentative conclusions. We have also included the results of
\cite{kantor} that consider a square well potential for the
``area" term (tethered surfaces).

After re-analyzing the results of \cite{npb} in the three-dimensional
case with the same binning criteria we have found plots for the
specific heat which are totally compatible with the ones reported
here for $S_2$. We tentatively conclude that $S_2$ and $S_3$ lead,
regardless of the form of the ``area" action, to transitions which are
most likely of higher order. This seems to agree with recent\cite{irback}
results
in dynamically triangulated surfaces (with a cosine action, however),
but leads to the distressing conclusion that there is some evidence
for non-universality.

\begin{table}
\setlength{\tabcolsep}{1.5pc}
\newlength{\digitwidth} \settowidth{\digitwidth}{\rm 0}
\catcode`?=\active \def?{\kern\digitwidth}
\caption{Tentative type of phase transitions}
\label{tab:types}
\begin{tabular}{lrrr}
\hline
		 & \multicolumn{1}{r}{$S_1$}
		 & \multicolumn{1}{r}{$S_2$}
		 & \multicolumn{1}{r}{$S_3$} \\
\hline
Area    & 2nd & 3rd & 3rd   \\
Gaussian & 2nd & 3rd &  ---  \\
Tethered  & 2nd & --- & ---  \\
\hline
\end{tabular}
\end{table}

\section{ACKNOWLEDGEMENTS}
We are grateful to J.Ambjorn, L.Jacobs, D.Johnston and J.F.Wheater for
discussions.


\begin{thebibliography}{9}
\bibitem{KPZdistKaz}  V.Knizhnik, A.Polyakov, A.Zamolodchikov, Mod. Phys.
		      Lett. A3 (1988) 819; J.Distler, H.Kawai, Nucl. Phys
		      B342 (1989) 342; E.Brezin, V.Kazakov, Phys. Lett.
		      236B (1990) 144.
\bibitem{polyakov}    A.Polyakov, Nucl. Phys. B268 (1986) 406
\bibitem{me}          F.Alonso, D.Espriu, Nucl. Phys. B283 (1987) 393
\bibitem{others}      L.Peliti, S.Leibler, Phys. Rev. Lett. 54 (1985)
1690;                       H.Kleinert, Phys. Lett. 174B (1986) 335
\bibitem{david}       F.David, E.Guitter, Europhys. Lett. 5 (1988) 709
\bibitem{dynamical}   J.Ambjorn, B.Durhuus, J.Frohlich, P.Orland, Nucl.
		      Phys. B270 (1986) 457; F.David, Nucl. Phys. B257
		      (1985) 543
\bibitem{tim}         T.Morris, Nucl. Phys. B341 (1990) 443
\bibitem{frohlich}    B.Durhuus, J.Frohlich, T.Jonsson, Nucl. Phys B240
		      (1984) 453
\bibitem{gross}       A.Billoire, D.Gross, E.Marinari, Phys. Lett. B139
		      (1984) 75
\bibitem{plb}         D.Espriu, Phys. Lett. 194B (1987) 271
\bibitem{npb}         M.Baig, D.Espriu, J.Wheater, Nucl. Phys. B314 (1989)
		      587
\bibitem{harnish}     R.Harnish, J.Wheater, Nucl. Phys. B350 (1991) 861
\bibitem{kantor}      Y.Kantor, D.Nelson, Phys. Rev. Lett. 58 (1987) 2774
\bibitem{irback}      A.Irback, Proc. of Lattice '91, North Holland 1992.
\bibitem{gaussian}    J.Ambjorn, B.Durhuus, J.Frohlich, Nucl. Phys B256
		      (1985) 433

\end{thebibliography}
\end{document}